\documentclass[aps,pra,floatfix,showpacs,groupedaddress,notitlepage,twocolumn,
amsmath,amssymb]{revtex4}

\usepackage{graphicx}
\usepackage{bm}
\usepackage{subfigure}
\usepackage{braket}
\usepackage{epstopdf}
\usepackage{times} 
\usepackage{color}
\usepackage{units} 
  
\renewcommand{\vec}{\mathbf}
\newcommand{\expect}[1]{\langle #1 \rangle}
\renewcommand{\b}[1]{\hat{b}_{#1}^{\color{white}\dagger\color{black}}}
\newcommand{\bd}[1]{\hat{b}_{#1}^{\dagger}}
\newcommand{\BHM}{\mathrm{BH}}
\renewcommand{\S}{\mathrm{S}}
\newcommand{\J}[2]{{\bar J}_\text{#2}^{\,\text{#1}}}
\newcommand{\fig}[1]{Fig.~\ref{#1}}

\begin{document}

\title{Cluster Gutzwiller method for bosonic lattice systems} 

\author{Dirk-S\"oren L\"uhmann}

\affiliation{Institut f\"ur Laser-Physik, Universit\"at Hamburg, Luruper Chaussee 149, 22761 Hamburg, Germany}

\textheight246mm
 

\begin{abstract}
A versatile and numerically inexpensive method is presented  allowing the accurate calculation of phase diagrams for bosonic lattice models. By treating clusters within the Gutzwiller theory, a surprisingly  good  description of quantum fluctuations beyond the mean-field theory is achieved approaching quantum Monte-Carlo predictions for large clusters. Applying this powerful method to the Bose-Hubbard model, we demonstrate that it yields precise results for the superfluid to Mott-insulator transition in square, honeycomb, and cubic lattices. Due to the exact treatment within a cluster, the method can be effortlessly adapted to more complicated Hamiltonians in the fast progressing field of optical lattice experiments.  This includes state- and site-dependent superlattices, large confined atomic systems and disordered potentials, as well as various types of extended Hubbard models.  Furthermore, the approach allows an excellent treatment of systems with arbitrary filling factors. We discuss the perspectives that allow for the computation of large, spatially-varying lattices, low-lying excitations, and time evolution.         
\end{abstract}


\maketitle

\enlargethispage{-5mm}

The recent progress in the realization of optical lattices for ultracold atoms offers an ideal testing ground for theoretical models and computational methods. Compared with their solid-state counterpart, the foremost advantage of the atomic systems  is the outstanding controllability  of interaction strength and lattice parameters. In particular, optical lattices allow to study Hubbard models for bosonic particles, where an important milestone was the observation of the superfluid (SF) to Mott insulator (MI) transition \cite{Greiner2002,Jaksch1998}. 
These new experimental possibilities have stimulated the development of theoretical methods for bosonic systems ranging from density-matrix renormalization group (DMRG) techniques and quantum Monte-Carlo (QMC) methods to mean-field theories. 

In mean-field theories, a single lattice site is decoupled from the surrounding lattice, where the fluctuations are described by a \textit{mean-field} parameter. This strong simplification allows nonetheless the qualitative description of strongly correlated phases such as the Mott insulator \cite{Fisher1989,Jaksch1998,Oosten2001}.  The decoupling is also achieved in the  Gutzwiller method \cite{Rokhsar1991,Krauth1992,Zwerger2003,Buonsante2008,Trefzger2011} where the wave function is expanded in local Fock states with individual coefficients.  Both limits in the phase diagram--the Mott state and the superfluid state--can be described (approximately) by  a product of local Gutzwiller states. In fact, it can be shown that perturbative mean-field theories and the Gutzwiller theory predict equivalent SF-MI transition points \cite{Krauth1992,Sheshadri1993}. The Gutzwiller method is very versatile and can  be  applied to various types  of  optical lattice systems ranging from homogeneous lattices to large spatially-varying lattices systems. It allows both to treat stationary states (e.g. Refs.~\cite{Krauth1992,Zwerger2003,Huber2008,Trefzger2011}) as well as to perform time evolution (e.g. Refs.~\cite{Jaksch2002,Damski2003,Trefzger2011,Bissbort2011,Fischer2011}). However, relying on the mean-field decoupling the Gutzwiller theory has  similar restrictions as other mean-field methods and the phase boundaries can only be calculated quantitatively.  

\begin{figure}[b]
\includegraphics[width=1\linewidth]{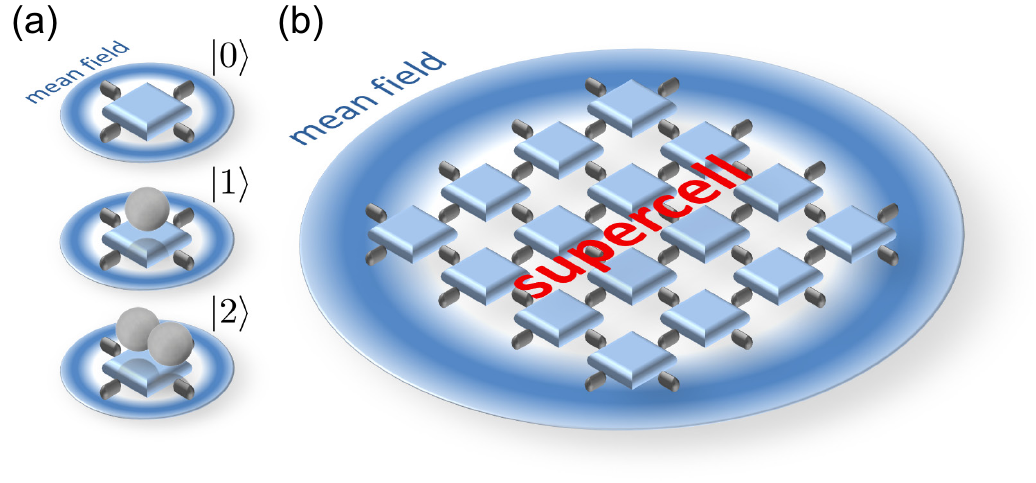}
\vspace{-20pt}
\caption{(a) Within the Gutzwiller theory a single lattice site is coupled to the mean-field $\expect{\b{j'}}$ of its nearest neighbors (illustrated for a square lattice). The lattice sites are expanded in Fock states $\ket{n}$ with $n=0,1,2,...$ particles. (b) In the supercell method, the single site is replaced by a cluster of lattice sites and is expanded in the many-site Fock basis $\ket{N}$. 
\label{Principle}
}
\end{figure}

Early, it was suggested by Bethe \cite{Bethe1935},  Peierls  \cite{Peierls1936},  and Weiss \cite{Weiss1948} that the mean-field method may be extended by coupling a cluster of sites (a \textit{supercell}) with the mean-field rather than a single lattice site. Usually, such a cluster is formed by a central site and its nearest neighbors. These cluster mean-field methods were mainly used for the description of electrons in solids \cite{Senechal2000,Senechal2002,Potthoff2003} and spin models \cite{Dantziger2002,Du2003,Etxebarria2004,Neto2006,Yamamoto2009} but also in the context of dynamical mean-field theories \cite{Kotliar2001}.
Recently, a multi-site mean-field method for the Bose-Hubbard model  was proposed in Refs.~\cite{McIntosh2012,Pisarski2011,Buonsante2004,Jain2004}  based on neglecting quadratic terms in the fluctuations.  However, due to the restriction to  relatively  small clusters quantum fluctuations could not be sufficiently included leading to inaccurate predictions of the phase boundary in the SF-MI phase diagram. Furthermore, bosonic cluster mean-field theory has been applied to dipolar hard-core bosons in Refs.~\cite{Yamamoto2012,Yamamoto2012b}.

Here, we present a cluster mean-field method for large clusters based on the well-established Gutzwiller method (\fig{Principle}). As a demonstration, this \textit{cluster} Gutzwiller method is applied to the Bose-Hubbard model in square, honeycomb, and cubic lattices. Using periodic boundaries  for the cluster,  it allows for accurate predictions  of  the SF-MI transition approaching quantum Monte-Carlo results for large supercells. We show that the approach is suited for the treatment of arbitrary filling factors and large-sized lattices. 
We start by reflecting several aspects of the single-site Gutzwiller method and a general description of the supercell method.
Subsequently, we apply the algorithm to the Bose-Hubbard model in several lattice geometries. Finally, details of the method are discussed that allow for the computation of large clusters. As an outlook, we discuss further applications such as the treatment of large lattices with non-equivalent sites, the determination of excited states or time evolution. 

\section{The Single-site Gutzwiller Approach}

Let us start with the standard Gutzwiller method which assumes that we can write the wave function of the system as a product of single-site wave functions. Each site $j$ is represented by the Gutzwiller trial wave function $\ket{j}=\sum_n c_n \ket{n}$ \cite{Rokhsar1991,Krauth1992,Zwerger2003,Buonsante2008,Trefzger2011} in the basis of local Fock states $\ket{n}$ with $n$ particles (\fig{Principle}a).
Usually the  coefficients $c_n$ are determined by imaginary time evolution \cite{Trefzger2011}. However, as detailed below also a self-consistent diagonalization scheme can be used to determine the coefficients. 
Very generally, the wave function of the whole system can be written as $\ket{j}\ket{\psi}$, where $\ket{\psi}$ is the wave function of all sites except for site $j$. Common tight-binding Hamiltonians can be written as 
\begin{equation} 
	\hat H=\hat H_\psi + \hat H_j + \hat H_{\psi j}  =\hat H_\psi + \hat H_j + {\sum}_\alpha \hat A_{\psi}^\alpha\, \hat B_{j}^\alpha ,
	 \label{eq:HPsiJ}
\end{equation}
where $\hat H_\psi$ and $\hat H_j$ act only on the respective subsystems.  
The last term represents the coupling between $\ket{\psi}$ and $\ket{j}$ and can be decomposed in a sum of subsystem operator products. 
As we show in the following, the initial knowledge of $\psi$ is not required when using a self-consistent loop for the site $j$.  
Assuming, however, that we know $\ket{\psi}$, the Hamiltonian matrix of the whole system in the Fock state basis $\{\ket{n}\}$ is given by
\begin{equation} \begin{split}
	H_{mn}=&\bra{\psi} \hat H_\psi \ket{\psi} \delta_{mn} + \bra{m} \hat H_j \ket{n} +\\
	& {\sum}_\alpha \bra{\psi} \hat A_\psi^\alpha \ket{\psi}    \bra{m} \hat B_j^\alpha \ket{n},
	\label{eq:Hmn}
\end{split}\end{equation}
where the first term is a constant energy offset. Now, the wave function on site $j$ can be obtained by diagonalizing $H_{mn}$ using a small Fock basis. 
For the Bose-Hubbard Hamiltonian 
\begin{equation}
	\hat H_\BHM= -J \sum_{\expect{j,j'}} \bd{j} \b{j'} + \frac{1}{2} U \sum_j \hat n_j (\hat n_j-1) - \mu \sum_j  \hat n_j,
	\label{eq:HubbardModel}
\end{equation}
with bosonic annihilation $(\b{j})$ and creation $(\bd{j})$ operators on site $j$, on-site interaction $U$, tunneling matrix element $J$, and chemical potential $\mu$, 
we obtain for the uncoupled part of the Hamiltonian 
$ 
	\bra{m} \hat H_j \ket{n}  =   \frac{1}{2} U\, n ( n-1)\, \delta_{mn} -\mu\, n\,  \delta_{mn}.
$ 
The coupling $\hat H_{\psi j}$ between $\ket{\psi}$ and $\ket{j}$  is given by
\begin{equation}
	\hat H_{\psi j}=-J\, \Big( \bd{j}  \sum_{\expect{j'}} \expect{\b{j'}} + \b{j}  \sum_{\expect{j'}} \expect{\bd{j'}} \Big) 
	\label{eq:HHopping}
\end{equation}
where the $\expect{j'}$ indicates the  summation over all nearest neighbors of site $j$. Using the Fock coefficients $c'_n$ of the trial wave function on site $j'$, we can obtain $\expect{\b{j'}}=\bra{\psi} \b{j'} \ket{\psi}=\sum_n {c'_n}^{\!\!*} c'_{n+1} \sqrt{n+1}$. 
At this point it is clear that the Gutzwiller method is equivalent to the mean field treatment   for the Bose-Hubbard model  as a single lattice site couples only to the average mean field $\expect{\b{j'}}$  and its conjugated.  In general, the Gutzwiller method has more internal degrees of freedom which becomes apparent, e.g., for occupation-dependent models \cite{Luhmann2012}. 
Note that the expression \eqref{eq:HHopping} allows for the calculation of large-sized inhomogeneous lattices by iteratively transversing through the lattice. For a homogeneous lattice with equivalent sites and $z$ nearest neighbors, the off-diagonal matrix elements of the coupling are given by
$	\bra{n+1} \hat H_{\psi j}  \ket{n} =-z \, J\,  \sqrt{n+1} \ \expect{\hat b} $
and its conjugated. 
By diagonalizing  $H_{mn}$, a new expectation value of the superfluid order parameter $\expect{\hat b}=\expect{\b{j}}$ for the site $j$ can be computed. Using a self-consistent loop, the Hamiltonian can be solved without initial knowledge of $\ket{\psi}$.

\section{The Cluster Gutzwiller Approach}

\begin{figure*}[t]
\includegraphics[width=0.95\linewidth]{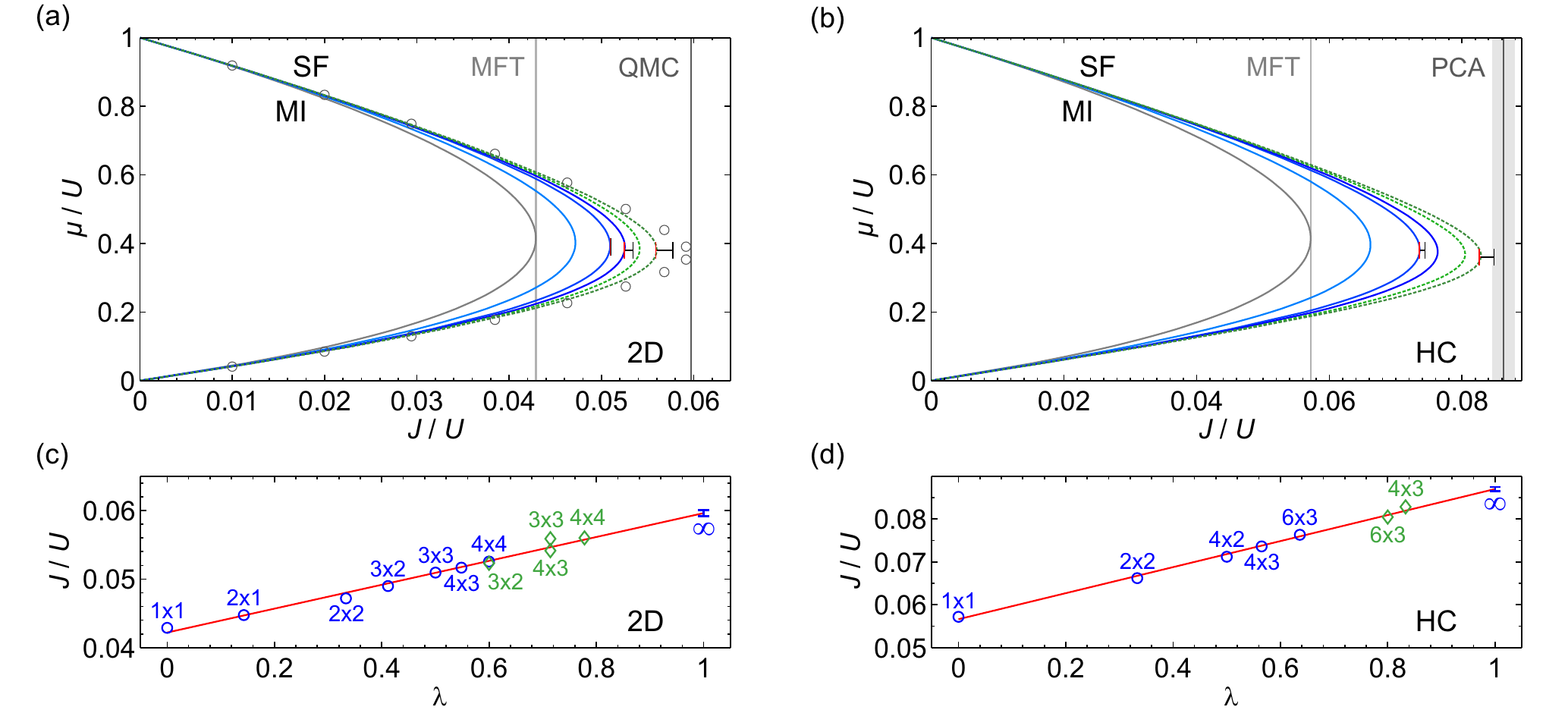}
\vspace{-8pt}
\caption{(a) The cluster Gutzwiller method applied to the Bose-Hubbard model of a square two-dimensional (2D) lattice. The gray lobe corresponds to the single-site Gutzwiller method  and  the blue lines to supercells with $s=$ 4, 9, and 16 sites (from left to right). Using periodic boundary conditions for $s=12$ and 16 sites sites (dashed green lines) improve the results significantly. The vertical lines depict the critical ratio $J/U$ obtained by mean-field theory (MFT) \cite{Fisher1989,Krauth1992,Oosten2001} and quantum Monte-Carlo (QMC). The QMC results (open circles) are taken from Ref.~\cite{Capogrosso2008}. The cluster calculation takes $f=7$ fluctuations into account (see text), where the red error bar corresponds to $f=\infty$ ($f=8$ for $s\geq16$) and the black error bar to $f=5$. 
(b) Results for the honeycomb (HC) lattice with three nearest neighbors for $s=$ 4, 12, and 18 sites (blue) as well as $s=$18 and 12 sites  with periodic boundary conditions (dashed green). 
The vertical line corresponds to the process chain approach (PCA) where the shaded area is the estimated error \cite{Teichmann2010}. 
(c,d) Finite-size scaling of the critical points for different cluster sizes, where circles (diamonds) indicate results without (with) periodic boundary conditions. The scaling parameter $\lambda$ is zero for a single site and one for an infinite lattice (see text).  For (c) 2D and (d) HC lattices, the predicted critical points for $\lambda=1$ match with the QMC and the PCA prediction, respectively, within the error bounds.   
\label{2DHC}
}
\end{figure*}

In the following, the cluster approach is described as an extension of the single-site Gutzwiller method.
For this it is important that the decoupling in equation \eqref{eq:Hmn} holds 
for an arbitrary subsystem of the lattice rather than a single lattice site. Let us therefore replace the single-site state $\ket{j}$ by a supercell cluster $\ket{\S}$ with $s$ sites (see \fig{Principle}b). The internal degrees of freedom within the supercell allow for quantum fluctuation that are not covered within the single-site mean-field approach. The generalized trial wave function for the cluster is given in a many-site Fock basis $\{\ket{N}\}=\{\ket{n_0,n_1,n_2,...}\}$ reflecting all possible distributions of $n_i=0,1,2,...$ particles on the sites $i$ of the supercell. In analogy to the expressions \eqref{eq:Hmn} and \eqref{eq:HHopping}, the respective Hamiltonian matrix  for the Bose-Hubbard model  is given by 
\begin{equation}\begin{split}
	\hat H_{MN}&= \bra{M} \hat H_\S \ket{N} +  {\sum}_\alpha \bra{\psi} \hat A_{\psi}^\alpha \ket{\psi}  \, \bra{M} \hat B_{\S}^\alpha \ket{N}  \\
    &=\bra{M} \hat H_\BHM^\S  - J \sum_{j\in \partial\S} \nu_j \Big( \bd{j} \expect{\hat b}  + \b{j}  \expect{\bd{}} \Big)  \ket{N} ,
	\label{eq:HamiltionanMatrix}
\end{split}\end{equation}
where the energy offset $\bra{\psi} \hat H_\psi \ket{\psi}\! \delta_{MN}$ has been omitted.
Here,  $H_\BHM^\S$ is the Bose-Hubbard Hamiltonian \eqref{eq:HubbardModel} where the summations are 
restricted to sites within the supercell $\S$. 
The second term describes the coupling of all sites at the boundary $\partial\S$ of the cluster with the mean-field $\expect{\hat b}$. The prefactor $\nu_j$ reflects the number of bonds to the mean-field (for \fig{Principle}b the corners of the square have $\nu_j=2$ and the other edge sites $\nu_j=1$). 
Note that for more complicated Hamiltonians, $\bra{\psi} \hat A_{\psi}^\alpha \ket{\psi} $ is not reducible to a single mean-field parameter $\expect{\hat b_i}$ and can contain other internal quantities such as $\expect{ \hat n_i \hat b_i}$ obtained from the coefficients $C_N$. In this case, the presented method would differ from the mean-field approach in Refs.~\cite{McIntosh2012,Pisarski2011}, where fluctuations of quadratic order are neglected.  
 In equation \eqref{eq:HamiltionanMatrix}, we  further  assume a homogeneous lattice where all boundary sites couple to the same mean-field $\expect{\hat b}$ but it can be easily expanded  to super lattices or finite lattice systems. By solving the eigenvalue problem \eqref{eq:HamiltionanMatrix}  of the cluster (cf. Refs.~\cite{McIntosh2012,Pisarski2011}),  we obtain the lowest eigenvector $\ket{S}= \sum_N C_N \ket{N}$, which is used  to calculate the mean field
\begin{equation}
	\expect{\hat b} =\bra{S} \b{t} \ket{S} = \sum_{M,N} C_M^* C_N \bra{M} \b{t} \ket{N}.
\end{equation}
self-consistently on a \textit{target} site $t$.
We choose the target site $t$ to be the most central site of the supercell. This ensures that the mean-field minimally couples to the target site and quantum fluctuation can be included as good as possible.   Note that the higher eigenvalues of the Hamiltonian \eqref{eq:HamiltionanMatrix} correspond to local excitations of the system. 

The feasibility of the cluster Gutzwiller method relies basically on three technical aspects, (i) an adequate restriction of the many particle basis $\{\ket{N}\}$, (ii) an effective diagonalization procedure, and (iii) an accurate and effective algorithm to determine the boundaries in the phase diagram.
Before discussing the aspects (i)-(iii), we demonstrate the accuracy of the cluster Gutzwiller approach by applying it to the Bose-Hubbard Hamiltonian \eqref{eq:HubbardModel}.

\section{Bose-Hubbard Phase Diagrams}

\begin{figure}[t]
\includegraphics[width=0.95\linewidth]{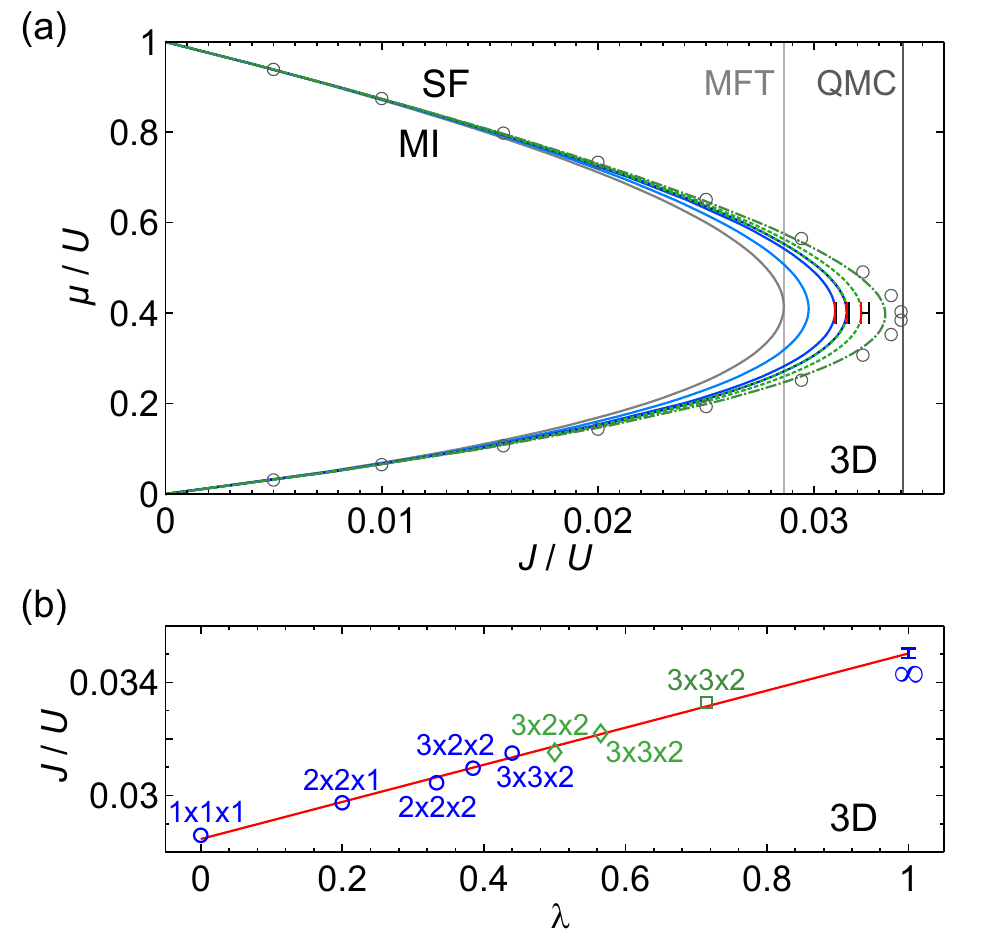}
\vspace{-8pt}
\caption{(a) Results for three-dimensional cubic lattices (3D) for supercells with $s=$ 4, 12, 18 sites (blue lines) and supercells with 12 and 18 sites using periodic boundaries (dashed green lines).  The blue and green line for 18 and 12 sites, respectively, nearly coincide. In three dimensions, periodic boundary conditions can be applied in two directions for the $3\!\times\!3\!\times\!2$ cluster (dotted-dashed green line). The QMC results  (open circles)  are taken from Ref.~\cite{Capogrosso2007}.  See \fig{2DHC} for further details. (b) The finite-size scaling for clusters without (circles) and with periodic boundary conditions in one (diamonds) and two directions (squares). 
\label{3D}
}
\end{figure}
 
For two-dimensional square (2D) and honeycomb (HC) lattices, the resulting phase diagrams for the SF-MI transition are shown in \fig{2DHC}.
The calculation for $s=1$ site corresponds to the Gutzwiller method and therefore is equivalent to the mean-field (MFT) result with the critical point $\J{}{MFT}=(J/U)_\text{crit}=1/(3+2\sqrt{2})z$ for the filling factor $n=1$ \cite{Fisher1989,Krauth1992,Oosten2001}. Here, the number of nearest neighbors is denoted by $z$, where $z=4$ for square and $z=3$ for honeycomb lattices. 
For the square lattice, the prediction $\J{2D}{MFT}=0.0429$ using mean-field theory differs substantially from the quantum Monte-Carlo method \cite{Capogrosso2008} and the strong coupling expansion \cite{Elstner1999} with $\J{2D}{QMC/SC}=0.0597$.
By increasing the number of sites $s$ in the cluster Gutzwiller method it is possible to correct the mean-field result substantially.  
While for an infinite number of sites $s$ the mean-field coupling at the boundary is negligible and the exact result is obtained, in practice, the cluster size is limited to $s=4\!\times\!4=16$ sites. Figure~\ref{2DHC} demonstrates that with an increasing number of sites the results are improving.   
Periodic boundary conditions along one spatial dimensional can improve results for finite-sized lattices significantly, since boundary effects are reduced. Already for a $3\!\times\!3$ cluster in 2D, where the computation is very inexpensive, the phase diagram is surprisingly accurate with $\J{2D}{ $ 3\!\times\!3 $ }=0.0559$ at the tip of the Mott lobe. 
Note that the absolute deviation from the numerical exact QMC results (open circles in \fig{2DHC}a) is the largest at the tip of the Mott lobe.    

For the accuracy of the cluster method the crucial factor is the ratio of internal bonds (within the supercell) to mean-field bonds at the boundary.
 By applying a scaling of the cell size as introduced in Ref.~\cite{Yamamoto2012b}, the critical $J/U$ for an infinite lattice can be interpolated. Here, the scaling with the parameter $\lambda=B_\S / (B_\S+B_{\partial\S})$ is performed and plotted \fig{2DHC}c, where $B_\S$ represents the number of bonds within the cluster and  $B_{\partial\S}$ the bonds to the mean field. The scaling parameter also explains why periodic boundary conditions (green diamonds) improve the numerical results. The predicted value $\J{2D}{CGW}=0.0596(4)$ for an infinite lattice ($\lambda=1$)  matches with the QMC data within the standard deviation.  

For the honeycomb lattice (\fig{2DHC}b), the reduced number of nearest-neighbors causes an even larger error of the mean-field theory  value   $\J{HC}{MFT}=0.0572$. 
The cluster method substantially corrects this single-site value and predicts $\J{2D}{CGW}=0.0870(5)$  for an infinite lattice. This coincides with the process chain approach in Ref.~\cite{Teichmann2010}  resulting in $\J{HC}{PCA}=0.0863$ within the estimated errors.  

For the three-dimensional cubic lattice (3D), the cluster method predicts $\J{3D}{CGW}=0.0350(2)$ (\fig{3D}). The QMC calculation gives a slightly lower value of $\J{3D}{QMC}=0.0341$ \cite{Capogrosso2007}. This deviation might be caused by the small edge lengths of the clusters in three dimensions. 

\begin{figure}[b]
\includegraphics[width=0.95\linewidth]{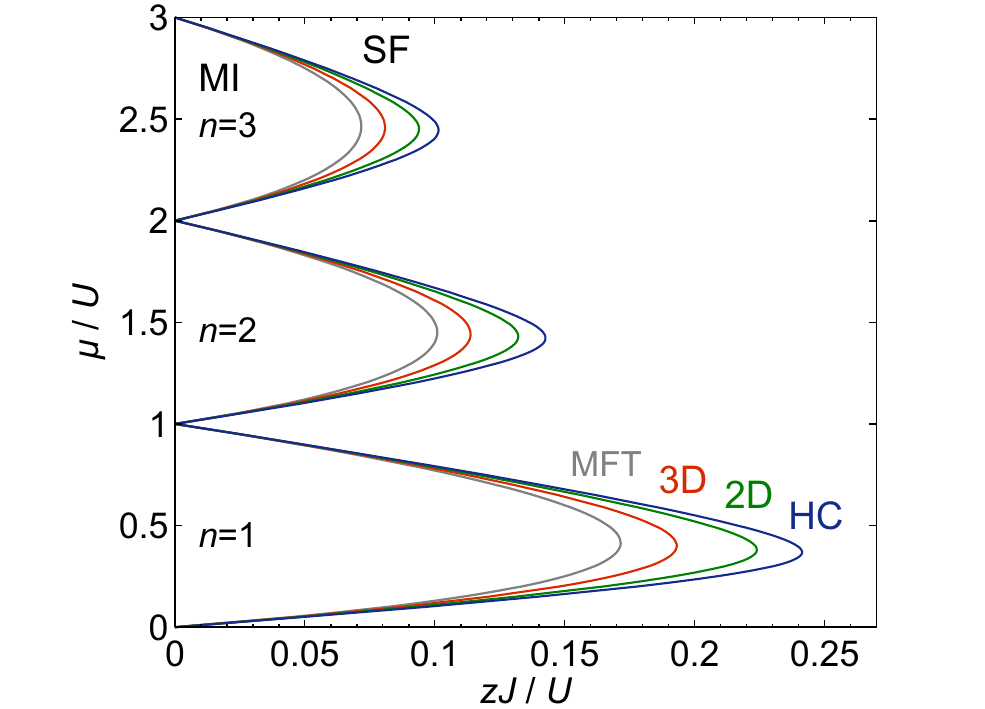}
\vspace{-8pt}
\caption{Mott insulator phase boundaries for filling factors $n=1$-$3$ obtained by the supercell approach, which allows for the calculation of arbitrary filling factors. The results for the cubic (3D), the square (2D), and the honeycomb lattice (HC) are plotted for the largest cell sizes with periodic boundary condition along one direction as a function of  $zJ/U$.
In this unit the mean-field boundary (gray) is independent of the lattice geometry. 
}
\label{MIs}
\end{figure}

The method presented here allows the treatment of systems with arbitrary filling factors $n$. The results for the lowest three Mott lobes are shown in \fig{MIs}. Note that the numerical effort does not increase with the filling factors, since only site occupations $n_i=n\!-\!2,...,n\!+\!2$ are of interest, which is elaborated in the next section.

\section{Treatment of large clusters}

We now turn back to the description of the numerical algorithm that allows to treat large cluster sizes. 
 The efficient implementation allows a numerical inexpensive computation of the phase diagram. For a given $\mu$ on a todays average (single-core) processor, the computation of $9$ sites in 2D ($5$ fluctuations) takes only few tenths of a second. Here, the method also allows to calculate the excitation spectrum. The computation time growths drastically with the cluster size, e.g. about $10$ seconds for $16$ sites ($\times15$ for $25$ sites and $\times4 $ for $6$ fluctuations).

(i) To apply the method described above, the infinite many-particle basis set has to be restricted to a finite but sufficient subset $\{\ket{N}\}$. The number of states in $\{\ket{N}\}$ grows exponentially with the particle number and therefore also with the number of cluster sites, which limits practically the size of the supercell. However, the complexity of the supercell problem can be drastically reduced depending on the specific Hamiltonian. From the single-site Gutzwiller theory for the Bose-Hubbard model, it is known that the MI state with $n$ particles is unstable at its boundary only to fluctuations with  $n\pm1$ particles; i.e., only zero, one and two particle Fock states have to be taken into account for the MI with a filling of $n=1$ \cite{Krauth1992}. This is, however, not completely true for the supercell method which has  internal degrees of freedom. It turns out that local fluctuations with $n_i=\pm2$ have an effect, if rather small, whereas higher local particle number fluctuations are extremely small and can also be neglected (see errorbars in Figs.~\ref{2DHC} and \ref{3D}).   However, we can indeed restrict the total particle number to $N$ and $N\pm1$.  
 Moreover, at the phase boundary also states with all sites simultaneously fluctuating are unlikely to be occupied. In practice, more than $f= \sum_i |n_i-n|\leq5$ fluctuations for smaller clusters and $f\leq7$  for larger clusters ($s\geq 16$) hardly change the results of the phase diagram (see error bars in \fig{2DHC} and \fig{3D}).   
 In the vicinity of the boundary the actual value of the superfluid order parameter is influenced by this constraints but not the criticality.
 
(ii) For the diagonalization of the coupled supercell problem Lanczos-based algorithms for sparse matrices can be used to obtain the lowest eigenvector \cite{ARPACK}. In fact, the structure of the Hamilton matrix has to be created only once for a given number of lattice sites. As usually consecutive points of a phase diagram are calculated, excellent guesses for both $\psi$ and the eigenvector can be provided speeding up the Lanczos diagonalization immensely.  
Furthermore, symmetries of the cluster can be used to build a symmetrized basis set, e.g., for a $4\!\times\!4$ cluster with periodic boundaries the $C_2$ and the $C_4$ symmetry allow a reduction of the basis length by a factor of $8$. While here we restrict ourselves to the ground state, this optimized method is also suited to compute a large number of excited states.

(iii) A straightforward implementation would calculate all points in the $\mu/U$ -- $J/U$ plane of the SF-MI phase diagram. In addition, the self-consistent loop convergences only slowly directly at boundary due to the sudden change of the superfluid order parameter $\varphi=\expect{\hat b}$, which would require many iterations to determine the phase boundary accurately. However, it is much more convenient to  rely only on the fact that the algorithm converges monotonically. Starting with an initial guess $\varphi_\text{th}$ (threshold value),  a \textit{single} iteration allows to determine whether the \textit{exact} value of $\varphi$ is smaller or greater than $\varphi_\text{th}$. As the value of $\varphi$ has a jump at the boundary, a small  but finite $\varphi_\text{th}$, such as $10^{-6}$, allows therefore to determine whether a given point is in the SF or the MI phase. Applying a binary search algorithm with only $20$ iterations for a given value of $\mu$ allows to determine the critical value with a relative precision of $2^{-20}\approx 10^{-6}$.  

\section{Conclusions}

We have presented an intuitive cluster mean-field method  based on the Gutzwiller theory where the single-site Fock state is replaced by a supercell. Using large clusters, we demonstrate for the Bose-Hubbard Hamiltonian that this method allows accurate results for various lattice geometries and arbitrary filling factors, in particular, for 2D cubic and honeycomb lattices. The approach can be adapted to various types of Hamiltonians and is numerical inexpensive. An intrinsic advantage of the method is that it can be easily applied to very large lattices with inequivalent lattice sites \cite{Pisarski2011}, such as disorder potentials or macroscopic confined  systems. To achieve this, the supercell  centered at site $\vec x$  can be iteratively moved through the lattice where each time the mean field  $\varphi_{\vec x}= \expect{\b{\vec x}}$  at the target site  $\vec x$  is updated. Another advantage is that the method can be used to compute the local excitation spectrum.   
Furthermore, it also allows to perform correlated time evolution.  
For each short time step $t_{i+1}$ and for each lattice site $\vec x_t$, the exact time-evolution can be performed within the supercell centered at site $\vec x_t$ using the time-dependent mean-field boundary $\varphi_{\vec x}(t_{i})$. This determines the expectation values $\varphi_{\vec x}(t_{i+1})$ for the next time step. 

\section{Acknowledgments}

I would like to thank O. J\"urgensen, K. Sengstock, and W. Hofstetter for stimulating discussions.

\end{document}